Variable frequency characterization of interaction at nanoscale in linear dynamic AFM:

an FFM primer.

Simon Carpentier[1], Mario S. Rodrigues[2] and Joël Chevrier[1,*]

1-Université Grenoble Alpes, Institut NEEL, F-38000 Grenoble, France and CNRS Institut NEEL, F-38042 Grenoble, France

2-Departamento de Física da Faculdade de Ciências da Universidade de Lisboa Campo Grande, Edifício C8 P-1749-016 LISBOA, Portugal (Dated: October 15, 2014)

* to whom email should be addressed

Using electrostatic coupling between an AFM tip and a metallic surface as a test interaction, we here present the measurement of the force between the tip and the surface, together with the measurement of the interaction stiffness and the associated dissipation. These three quantities constitute a full characterization of the interaction at nanoscale. They are measured independently, simultaneously and quantitatively at the same place. This is made possible thanks to a force feedback method that ensures the DC immobility of the tip and to the simultaneous application of a sub-nanometer oscillation to the tip. In this established linear regime, stiffness and damping are directly obtained from amplitude and phase change measurements. The needed information for this linear transformation is solely the lever properties in the experimental context. Knowledge of k, its stiffness, $\gamma$, its damping coefficient and $\omega_0$, its first resonance frequency is shown to be sufficient in the frequency range we are here exploring.

Finally, we demonstrate that this method is not restricted to the lever resonance frequency. To the contrary, this interaction characterization whose resolution is limited by the Brownian motion, can be used at any frequencies with essentially the same performances. We believe that simultaneous and independent measurements of force, stiffness and damping, out of lever resonance, at nanoscale, and within the context of linear response define a new AFM paradigm that we call Force Feedback Microscopy (FFM). This article details the use of FFM using a well known and easy to implement electrostatic interaction between a regular AFM tip and a metallic surface in air. A graphical abstract is proposed at the very end of the article.

## I. INTRODUCTION

Modern Atomic Force Microscopes (AFM) enables one to manipulate AFM nanotip at nanoscale with extreme efficiency. Performances in the control of X, Y and Z displacements (far below the atomic scale), immunity against external vibrations, detection of the lever deformation at any frequencies and thus easily in the picometer range, and the associated signal over noise ratio limited by the Brownian motion, the ultimate thermodynamical limit, make these machines in essentially all key environments (vacuum, gas, liquids) a major tool to probe interaction at nanoscale. As shown long ago by Surface Force Apparatus (SFA) in the context of new born nanosciences [1], a very efficient measurement strategy relies on the ability to simultaneously and quantitatively measure the force between the probe and the surface, and the associated interaction stiffness and damping. In SFA, this is done through the direct measurement of the DC force and thanks to the linear machine response to an excitation, at different frequencies. Through a machine calibration, in this linear regime, force, stiffness and damping are quantitatively and simultaneously measured at different frequencies by SFA [2]. In this paper, following our previous studies [3,4,5,6], we want to further demonstrate that an AFM used as Force Feedback Microscope (FFM) can be transformed in a nano-SFA, that is a machine used to fully characterize the interaction at nanoscale in a linear response measurement strategy. Nano-SFA refers to the fact that a characteristic of the SFA is the lateral size of the probe, which is around a millimeter, which precludes local investigation of the interaction at nanoscale. Here, when FFM measurement scheme is used, the lateral size of the probe is in the nanometer scale as defined by the AFM tip.

Linear response scheme is certainly one of the most widely used measurement strategy in science. This is apparent in many fields such as information theory, physics and engineering. Different names exist to name the associated linear response functions such as susceptibility, dielectric constant, impulse response, impedance… Theoretical tools such as Green function methods have a central place in associated formal descriptions. The damped harmonic oscillator is often the chosen example to introduce linear response theory because of its manifestations and uses in most of these mentioned fields. It is also the most basic and robust tool to model an Atomic Force Microscope. However it is remarkable that it is not the advanced operating scheme proposed for AFM. This is clearly stated in [7]: « *…the motion of the cantilever is highly nonlinear, and in conventional dynamic force microscopy, information about the sample that is encoded in the deflection at frequencies other than the excitation frequency is irreversibly lost.* » AFM dynamic modes which essentially implies lever excitation at the resonance frequency, is highly non linear and requires implementation of a complex multifrequency analysis so that the interaction at nanoscale can be fully analyzed [8].

That AFM has to specifically face this sophisticated and complex non linear measurement scheme appears to be related to two characteristic features: i) the use of an AFM lever with a low stiffness to improve sensitivity in the measurement of lever deformation leads to a mechanical instability refereed as "jump to contact". This brutal and irreversible move of the tip toward the surface occurs as the force gradient related to the tip surface interaction becomes larger than the lever stiffness. This is also named "pull in" effect in the context of MEMS. Existence of this "jump to contact" usually prevents static measurement of the force in the most part of the attractive interaction curve [5]. A way to avoid this jump to contact when entering the close vicinity of the sample with the tip probe is to oscillate the lever. This oscillation changes the stability conditions of the lever and the conditions for "jump to contact" occurrences. One can here notice that in the current AFM literature, static force, especially in the attractive regime, is not usually directly measured. This is one of our claim here when using FFM; ii) further this periodic excitation of the lever enables the AFM user to probe the interaction variations at the scale of the amplitude of oscillation, and in principle to access the interaction stiffness and the characteristics of the associated energy dissipation. For stability purpose as mentioned in i) and to have an easier detection of the lever behavior, amplitudes of oscillations used are commonly between 1nm and 10 nm as stated in [7]. These oscillations are most often not small compared to the typical spatial variations of most interactions at nanoscale and this is essentially the source of non linearity.

When measuring forces, especially if the lever Brownian motion is not the limiting factor, it is of high interest to excite the lever at an eigenmode [9]. In case of a high quality factor as it is usual in air and vacuum, one immediately take the benefit of an immediate mechanical amplification of interaction effect on lever behavior. The same amplification clearly applies to the Brownian motion. When this quality factor is low as in liquids, and depending on the noise characteristics acting on the measurement set up, this advantage can become irrelevant. This aspect is further experimentally explored in this article.

Using excitation frequencies of AFM lever eigenmodes introduces a clear limitation: there is no a priori reason why lever eigenmode frequencies should be relevant frequencies for the system studied. In cases of soft condensed matter and biology, it is very likely that the system studied will have its own dynamics and characteristic times. There is however no other way to match the probe frequency to the system dynamics when using eigenmodes, than changing the probe, i.e. the AFM lever. In other words, it would be of high interest to be able to vary the used lever excitation frequency when probing the interaction at nanoscale. In a forthcoming research article, we shall report on the use of this strategy here introduced to probe the capillary bridge visco-elastic properties at different frequencies (from 300Hz to 100000 Hz) using a single lever.

The present article is the description of the FFM experimental method in the simplest possible case. FFM method is based on an immobile tip at low frequencies (DC behavior) and the linear response when dynamical and static behaviors are simultaneously probed. For the sake of the present experimental demonstration, we here use an electrostatic coupling between a tip and a surface. Following previous investigations of FFM modes related to tip stability and application to biological samples [3-6], we here show that the tip surface interaction can be fully characterized in linear regime, through simultaneous and independent measurements of the force and the force gradient versus the distance. Specifically we show that the interaction stiffness can be quantitatively measured at different frequencies (here at frequencies below, at and above the first eigenmode) and the numerical derivative of the measured static force can be compared to the interaction stiffness measured at different frequencies. That the electrostatic interaction is not a dissipative force is here an experimental result within the measure sensibility.

2. Experimental method and results:

   The cantilever is the same lever in all presented experiments. It is a BudgetSensors All In One-Al, with a stiffness 0.67 N/m, as determined from thermal fluctuations. Its measured resonant frequency is at 15.042 kHz and its quality factor in air is 43. The metallic surface is a copper film. We have performed Z-spectroscopy measurements in presence of a 10V electrostatic potential between the tip and the metallic surface. Maximum stiffness here measured is well below the lever stiffness. It means that no "jump to contact" would here occur in absence of the feedback loop, which keeps the tip immobile in space. The suppression of the "jump to contact" was addressed in previous articles [3-5].

a) Measurement of the static force using the force feedback method

   In Force Feedback Microscopy, the tip is at a fixed point in space. As the sample is moved closer to the surface, which increases the force on the tip, the piezoelement at clamped extremity of the lever, is moved so that the tip remains still. In order to achieve this, the total force acting on the tip must be equal to zero at any time. The regular AFM methods measure a lever deformation to finally acquire information about the tip surface interaction. Here we keep the tip position constant in space by opposing in real time the force applied by the surface to the tip. A parallel can be drawn with the two classical methods used to measure weight, and then masses, at our scale. Dynamometers are based on the measure of a spring deformation whereas Roberval balance aims at no displacement of plates by adding calibrated masses on the second plate.

   In order to achieve this as described in [3, 4], the tip position is constantly measured using an interferometer formed by two mirrors [9]. One is the end part of a cleaved optical fiber and the second is the lever back surface. The red laser beam injected into the optical fiber is partly reflected by these two surfaces and produces an interferometric signal highly sensitive to the distance between these surfaces, and therefore to the tip position. The bandwidth used to measure the tip position is larger than 0.5MHz. In order to keep the lever deformation at zero, a PID feedback loop drives the piezoelement at the clamped lever end. Throughout the whole experiment, this results in a tip position fixed in space with a precision better than 0.1nm. The bandwidth used to act on the tip position so that it does not move can be up to 100kHz [4]. The measured static force is then simply the product of the lever stiffness and the piezoelement controlled displacement of the clamped lever end.

For an applied voltage of 10 volt between tip and surface, the static force determined using this method as the tip-surface distance is varied is shown in figure 1:

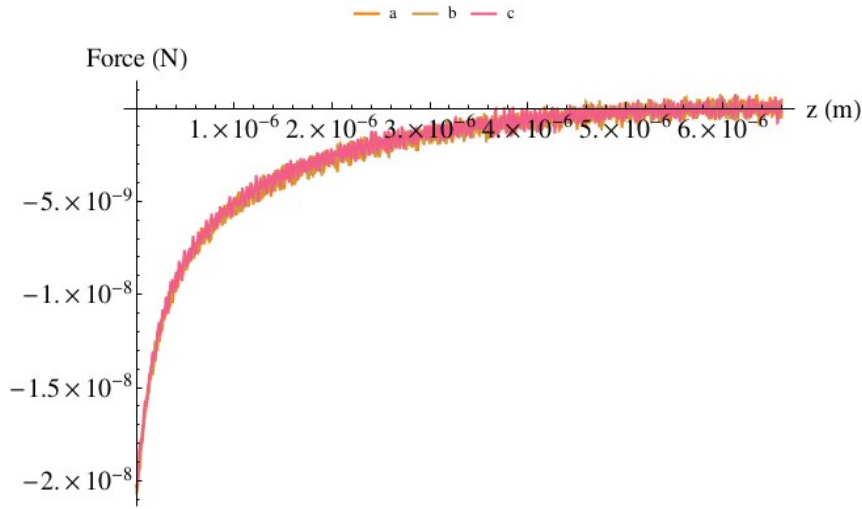

Figure 1: Measured static forces between the tip and the metallic surface at different tip surface distances are presented. The zero on distance axis is here in an arbitrary position. A zero

determination will be proposed at the paper end. Three curves are superimposed. They are associated to dynamical measurements at 10, 15 and 20kHz: at each frequency, the static and dynamical curves have been measured simultaneously.

Our absolute precision on the force measurement is essentially determined by the uncertainty on the value of the lever spring constant, as it is the case for all AFM methods.

b) Measurement of the force gradient at different frequencies

In all subsequent described experiments, the static position of the tip is kept constant in space using the method described in the previous section of this paper [3]. Measured experimental parameters are the tip oscillation amplitude and the phase shift between the tip oscillation and the applied excitation. From these two measured independent parameters, the interaction stiffness and the damping coefficient in N/m are determined using a linear 2x2 transformation.

From [4], we have the correspondence between the damping coefficient $\gamma_m$, i.e. the imaginary part of the system response, to the applied periodic excitation, and the real part that is the interaction stiffness km. The interaction stiffness is here directly the negative of the force gradient as shown experimentally below.

$$k_m = F_r [n \cos(\Phi) - \cos(\Phi_\infty)] \quad (1)$$

$$\gamma_m = F_r [-n \sin(\Phi) + \sin(\Phi_\infty)] \quad (2)$$

n and $\Phi$ are the measured quantities as tip surface distance is varied. $\Phi$ is the phase between excitation at the lever clamped end and the tip oscillation. n is the normalized amplitude of tip vibration i.e. the measured amplitude at distance d divided by the measured amplitude when tip is far away from the surface. At large distance n=1. The amplitude chosen here is all cases at large distances is 1nm. In this case here designed essentially to illustrate how FFM can be operated, such amplitude is small enough to ensure linear regime conditions in all presented measurements. Values of Fr and $\Phi\infty$ are directly obtained from the system transfer function when the tip is far from the surface but in the experimental environment used.

$$Fr = [(k-m\omega^2)^2 + \gamma^2\omega^2]^{1/2} \text{ and } \Phi\infty = \arctan[(\gamma/m)\omega/(\omega^2 - \omega_0^2)].$$

We here use 3 excitation frequencies. The first is below (10kHz), the second at (15kHz) and the third (20kHz) above the first lever eigenmode. In these three cases, a single mode description of the lever dynamics is sufficient, k, is obtained from measurement of the Brownian motion. This measure of k determines the precision of the force measurement. The damping $\gamma$ and the resonance pulsation $\omega_0$ (and therefore m) are obtained from lever transfer function measurement when the tip is far enough from the surface so that no tip surface interaction remains. Fr and $\Phi\infty$ determined using this method are directly used in equations (1) and (2) to obtain $k_m$ and $\gamma_m$ from n and $\Phi$.

The validity of this method has been here tested using an electrostatic interaction between a tip and a surface. As mentioned, different experiments were performed with lever excitation at 10 kHz (below the resonance frequency), 15 kHz (the resonance frequency) and 20 kHz (above the resonance frequency), but in all three measurements the final results, i.e. $k_m$ (d) and $\gamma_m$ (d) are supposed to be the same. $\gamma_m$ (d) should be essentially the noise floor and the three curves $k_m$ (d) are expected to be identical and to meet the derivative of the static force (in this frequency range, no characteristic time is here expected in the charge dynamic).

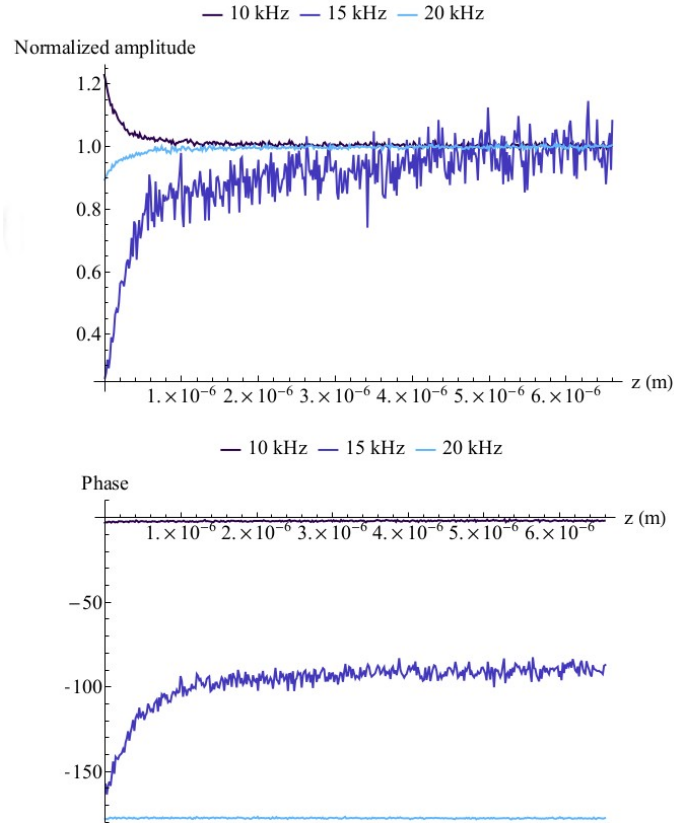

Figure 2: Normalized amplitudes and phase signals (as defined in text) measured at 10 kHz, 15 kHz and 20 kHz versus tip sample distance. An electrostatic potential difference of 10 V is applied between the doped silicon tip and the copper sample. Despite their large apparent differences, these three signals are measurements of the same interaction (see text and next figure).

The amplitude n (d) and phase Φ (d) for the three frequencies are presented in the Figure 2. The first striking observation is that the three signals are very different although they are three measures of exactly the same interaction. Beside, the amplitude and phase signals at 15 kHz, which is the resonant frequency, clearly present a noise level far larger than the two others. This is no surprise. The noise in this curve is essentially due to the Brownian motion at room temperature. The associated Langevin force applies equally to the lever at all frequencies. The induced displacement is essentially multiplied by 1/k at low frequency whereas Q/k multiplies it at the resonant frequency, which introduces a far larger noise at resonance. Using equations (1) and (2) will apply to these data, a factor Fr which is highly frequency dependent and will have the reverse effect. Multiplying amplitude and phase linear combination in equations (1) and (2) by k at low frequency and by k/Q at resonance frequency results in essentially the same noise signal ratio whatever the excitation frequency.

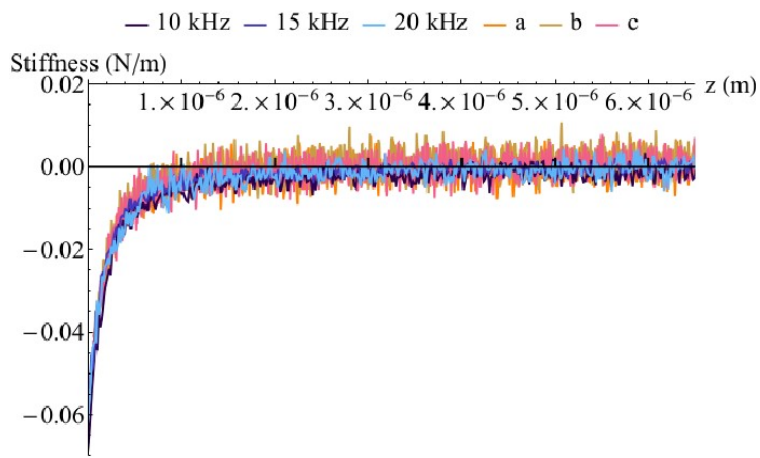

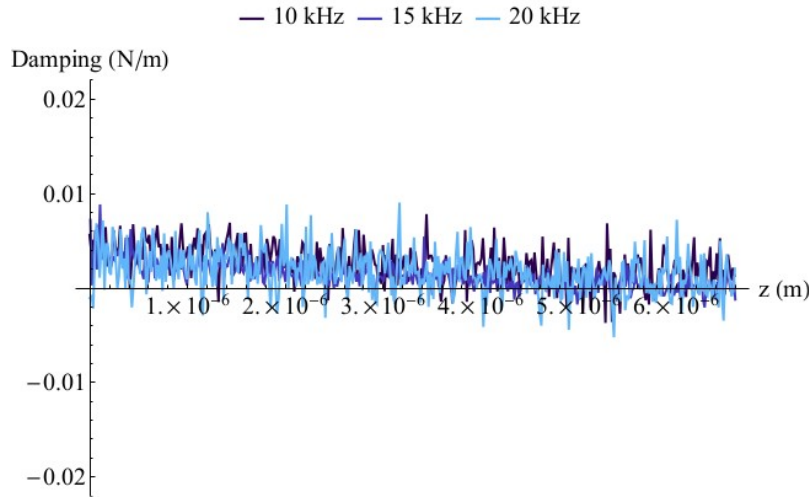

Figure 3.A) The interaction stiffness obtained with equation (1) are compared to minus the derivative of the static force [a), b) and c) are curves respectively measured at 10kHz, 15kHz and 20kHz].

3.B) Dissipation obtained for 10 kHz, 15 kHz and 20 kHz. Data obtained using equation (2).

In Figure 3.A, we present the stiffness obtained from the phase and amplitude signals for the three used frequencies. They are obtained applying equation (1) and using data from figure 2. The stiffness obtained are furthermore compared to minus the derivative of the static force. We first observe that although original data, amplitude and phase, strongly differs, the three km (d) curves are within experimental incertitude, identical at all distances. They also exhibit essentially the same noise level. This experimentally shows that, first one can perform experiment with an AFM out of resonance frequencies especially when it is operated in the linear regime of FFM, and second, that there is here no benefit when working at resonance frequency. The numerical derivative of the measured static force nicely matches the three-measured interaction stiffness.

This is a comparison of simultaneous but independent measurements as the only parameter in common between static and dynamic measurements is k the lever spring constant. We also observed that the noise on the calculated force gradient is far more important than on the measured stiffness. The static measurement is a broadband measurement and takes into account noise over a large frequency window. Further, as well known, additional 1/f noise coming from environment is adding at very low frequencies. All this result is an increased noise in static curve. This is not surprising.

Damping coefficient $\gamma_m$ (d) is reported in figure 3.B. As expected, the observed signals, at all three frequencies, are essentially noise. In the present experimental conditions, we do not expect any dissipation associated to the electrostatic interaction that is a conservative force. We have no grounded explanation for the slight observed increase of $\gamma_m$ (d) as distance is decreased. Following [10,11], we have tried to estimate a dissipation increase due to the vicinity of the surface. Approximate calculated values obtained do not provide reasonable figures for what is observed.

To further characterize the interaction, we have plotted inverses of static forces (1/F) measured at different tip surface voltages. We reports results in figure 4 together with forces versus distance in inset. The result here is that in all cases for this distance range, the force is proportional to $d^{-1}$. Such a dependence of the electrostatic force versus distance is typical of a sphere plane geometry [9]. The extracted value of an equivalent sphere radius is 1.3 micrometer. Not surprisingly as electrostatic interaction is a long range interaction, this is far larger than tip apex radius. The fact that the force variation versus distance is very close to $d^{-1}$ leads to the definition of an effective tip radius which is consistent with the tip structure.

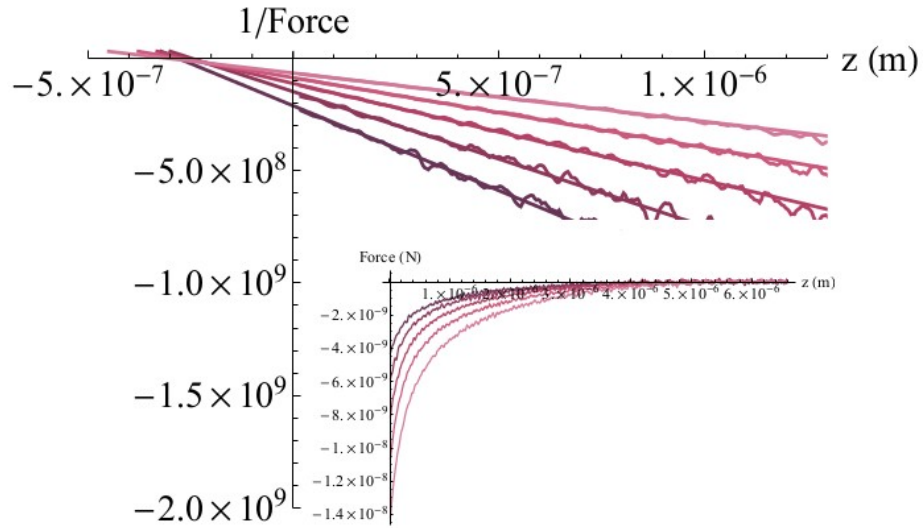

Figure 3. Inverses of the forces with respect to the tip-sample distance for tip-sample potentials varied from 6 V (purple) to 10 V, (pink). All linear extrapolations of the measured 1/F curves cross the distance axis at the same point, which define the electrical surface position and the electrical zero.

In inset : Measured static forces versus distances for different tip-sample potentials from 6 V (purple) to 10 V (pink)

In conclusion, we have shown that thanks to the linear response of the cantilever to a weak periodic excitation, and to a close loop setup that maintains the tip at a fixed position in space, we are able to trace back the full tip-sample interaction at arbitrary frequency and without any other calibration than the knowledge of the lever properties in the experimental context. In this operating scheme, we simultaneously and independently record the static force, the stiffness and the damping coefficient versus the distance.

This article essentially shows how to use an FFM to obtain a complete experimental characterization of the tip surface interaction. The choice of the electrostatic interaction is here to provide a clear experimental situation in order to demonstrate and to validate the FFM method. This method opens new possibilities to probe interaction at nanoscale especially in liquid when it is to explore visco-elastic properties in soft condensed matter or in biology as we have started to show [6]. Following SFA performances, the aim is to obtain quantitative measurements in probing tip surface exploration. FFM should be seen as a nano-SFA. In a coming article using exactly the experimentally method here shown, we shall show how it is possible to explore visco-elastic properties of the capillary bridge between a tip and a surface on a large range of frequencies (essentially from 300Hz up to 150kHz) in a single experiment using therefore a single lever.

Acknowledgments:

The authors thank Elisabeth Charlaix, Fabio Comin, Luca Costa and Peter van Zwol for enlightening discussions.

Graphical Abstract

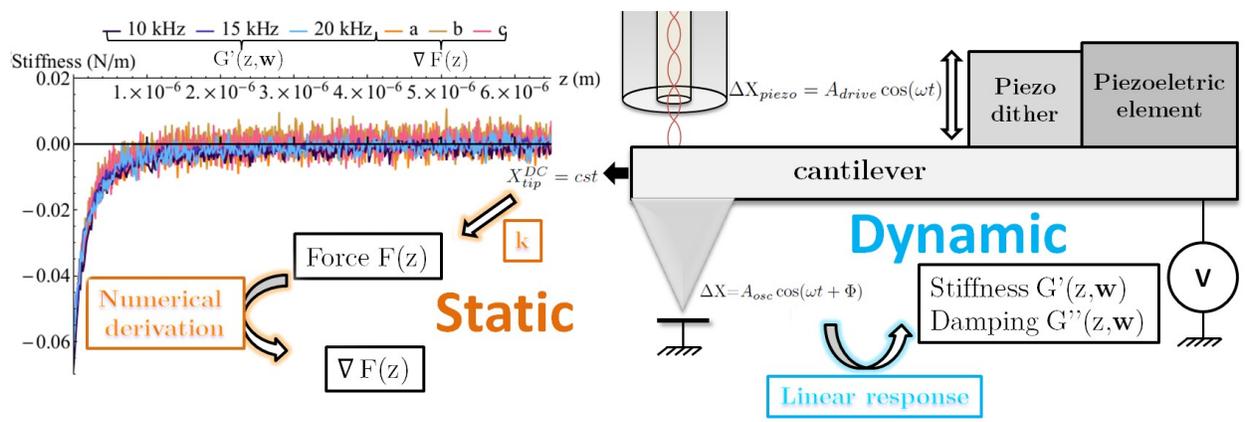